\documentclass[12pt]{article}
\usepackage{pazh}
\usepackage[english]{babel}
\usepackage{amsmath}
\usepackage{graphicx}
\usepackage{xcolor}
\usepackage{caption}

\graphicspath{{Pictures/}}

\tightenlines

\sloppypar

\begin{document}
	
\captionsetup[figure]{labelfont={bf},labelformat={default},labelsep=period,name={Figure}}
	
	\title{\bf Effects of dispersion of the dust velocity in the LISM on the interstellar dust distribution inside the heliosphere}

    \author{ E.A.Godenko\affilmark{1,2,3*}, V.V.Izmodenov\affilmark{1,2,3}}

    \affil{
        {\it Institute for Problems in Mechanics RAS, 119526, Moscow, Pr.Vernadskogo, 101-1}$^1$ \\
	    {\it Lomonosov MSU, Moscow Center for Fundamental and Applied Mathematics, 119992, Moscow, GSP-1 Leninskie Gory}$^2$ \\
        {\it Space Research Institute RAS, 117997, Moscow, Profsoyuznaya Str 84/32}$^3$}

    \sloppypar 
    \vspace{2mm}
    \noindent
    {\bf Annotation} -- Interstellar dust (ISD) penetrates into the heliosphere due to the relative motion of the Sun and the local interstellar medium (LISM). Inside the heliosphere and at the boundaries, where solar wind interacts with the LISM, distribution of ISD is modified due to the action of the electromagnetic forces, the solar gravitation and the radiation pressure. These forces make the distribution of the ISD particles in the heliosphere inhomogeneous. In previous work we demonstrated the existence of singularities in the ISD density distribution at 0.03 - 10 a.u. north and south with respect to the heliospheric current sheet. In this paper we show that dispersion in the ISD velocity distribution strongly affects the singularities. Even small values of dispersion have the drastic impact on the density distribution and smooth the high density layers discovered previously.


    \vfill
    \noindent\rule{8cm}{1pt}\\
    {$^*$ E-mail $<$eg24@yandex.ru$>$}
     
    \clearpage
    
    \section*{INTRODUCTION}
    
    \noindent
    The local interstellar medium (LISM) moves relative to the Sun with the speed $\sim 26 \:\:$km/s (Witte 2004, McComas 2015). Besides the plasma and neutral components, the LISM also contains dust component (Mann 2010). Unlike the plasma particles, the neutral and dust particles can penetrate into the heliosphere due to the relative motion. For example, the mean free path of neutral hydrogen due to charge exchange is $\sim 50-100 \:\:$a.u. (Izmodenov et al. 2000), comparable with the characteristic size of the heliosphere.
    
    \noindent
    The first evidence for the existence of interstellar particles in the heliosphere was the Lyman-$\alpha$ emission of interstellar neutrals (Bertaux, Blamont 1971). Direct measurements of the interstellar helium atoms were obtained by the Ulysses/GAS instrument (Witte 1992), and since the mean free path of interstellar helium is much larger than the size of the heliosphere, one can derive the macroscopic parameters of the LISM from these measurements. Nowadays, direct measurements of interstellar neutrals (hydrogen, oxygen and helium) are performed on IBEX using the IBEX-Lo instrument (e.g. Moebius et al. 2009, Katushkina et al. 2015, Baliukin et al. 2017). On the spacecraft SOHO (SWAN instrument) measurements of intensity and spectral characteristics of the Lyman-$\alpha$ emission are continuing (e.g. Qu{\'e}merais et al. 2013). Various models of the heliosphere are employed for the analysis of the experimental data (e.g. Izmodenov, Alexashov 2015, 2020, Pogorelov et al. 2011, Zirnstein et al. 2016).
    
    \noindent
    The ISD grains are solid grains with characteristic sizes in the range of hundreds of nanometers to microns (Mathis et al. 1977). Chemical composition of ISD is carbonaceous materials and astronomical silicates (Draine 2009). The mass fraction of ISD in the LISM is about 1 \% (Mann 2010). The ISD grains are charged positively as net effect of different physical processes such as photoelectron and secondary electron emissions. The presence of nonzero electric charge makes the trajectories of ISD more complex than of interstellar neutrals (not taking charge exchange with protons into account).
    
    \noindent
    It is difficult to detect ISD in the heliopshere because of the presence of interplanetary dust, which is emitted from asteroids, comets and other large objects in the Solar system. It is generally supposed that in the undisturbed LISM the interstellar dust is comoving with other components. This assumption was used in order to detect the ISD grains on Ulysses (Gr\"un et al. 1994). Moreover, the trajectory of Ulysses went significantly out of the ecliptic plane and thus it gave an opportunity to relatively easily separate interstellar dust from interplanetary dust, which is located principally in the ecliptic plane (e.g. zodiacal dust). Presence of ISD was also confirmed in the measurements on board the Galileo (Altobelli et al. 2005) and Cassini (Altobelli et al. 2007) spacecraft.
    
    \noindent
    The first models of the ISD distribution in the heliosphere were made by Bertaux, Blamont (1976) and Levy, Jokipii (1976). They studied the distinct influence of the gravitational and electromagnetic forces on the motion of the dust particles in the heliosphere. The next wave of interest in the ISD studying was associated with Ulysses measurements.
    Landgraf et al. (2000, 2003) analyzed these measurements using the Monte-Carlo modeling. They considered the combined influence of the gravitational, radiation pressure and electromagnetic forces on the particles in presence of time-dependent solar magnetic field. The ISD distribution and filtration of dust grains by the magnetic field at the heliospheric boundaries were explored by Czechowski, Mann (2003), Alexashov et al. (2016). Slavin et al. (2012) have built a 3D model of the ISD distribution for two opposite phases of the heliospheric magnetic field (focusing and defocusing). It is also taken account of the turbulence of the interstellar magnetic field and dependence of the surface charge potential on the heliocentric distance. Nowadays, the Monte-Carlo method is often used for theoretical studies of ISD. The descriptions and results of the modeling are shown in Sterken at al. (2012, 2019), Strub et al. (2015, 2019). These models are developed from the earlier model of Landgraf et al. (2000) using advanced numerical techniques and taking into account of the newer measurements. Mishchenko et al. (2020) applied a Lagrangian method (see Osiptsov 2000) to discover singularities in the distribution of ISD in the heliosphere. In the simplified stationary case when the heliospheric current sheet is a plane coinciding with the solar equatorial plane they demonstrated the existence of density singularities where the number density is infinite. They showed that the singularities form several dense dust layers for each size of the ISD particles on both sides of the current sheet. These singularities have never been observed in the previous papers studying dust distribution in the heliosphere by Monte-Carlo modeling because it requires a computational grid with an extremely high spatial resolution. In this paper we use a computational grid with cell size of $10^{-3} \:\:$a.u. and for studying local effects near density peculiarities --- of $10^{-6} \:\:$a.u.
    
    \noindent
    Mishchenko et al. (2020) used the assumption that the ISD particles have identical velocities in the LISM. Due to the fact that the ISD particles have nonzero electric charge, they interact with the interstellar magnetic field. Fluctuations of the  magnetic field lead to the acceleration of the charged dust particles (Hoang et al. 2012), that breaks the uniformity in the ISD velocity distribution in the LISM and adds some rather small dispersion. The goal of this paper is to study the influence of dispersion in the velocity distribution of ISD in the LISM on the emergence of the density singularities in the heliosphere. Slavin et al. (2012) also explore dispersion in the undisturbed LISM, but they do not study its influence on the singularities, since the computational grid used is quite coarse (5 a.u. for each direction).

    \section*{DESCRIPTION OF THE MODEL}
    
    \subsection*{Mathematical formulation of the problem}
    
    \noindent
    For the description of the ISD motion in the heliosphere we use a kinetic approach. In this way we should calculate the ISD distribution function $f_d(t, {\bf r}, {\bf v})$. The kinetic equation for $f_d(t, {\bf r}, {\bf v})$ is:
    
    \begin{equation} \label{formula: kinetic_equation}
        \frac{\partial f_d}{\partial t} + {\bf v} \cdot \frac{\partial f_d}{\partial {\bf r}} + {\bf F} \cdot \frac{\partial f_d}{\partial {\bf v}} = 0,
    \end{equation}
    
    \noindent
    where ${\bf F}$ is the sum of forces acting on the dust particles. On the right hand side of (\ref{formula: kinetic_equation}) we have zero, because in the heliosphere one can neglect collisions between dust grains and their interaction with plasma protons and electrons (Gustafson 1994). In this article we consider a stationary model of the solar magnetic field in the focusing phase, that is why the solution of the kinetic equation is also stationary, $\frac{\partial f_d}{\partial t} = 0$:
    
    \begin{equation} \label{formula: simplified_kinetic_equation}
        {\bf v} \cdot \frac{\partial f_d}{\partial {\bf r}} + {\bf F} \cdot \frac{\partial f_d}{\partial {\bf v}} = 0.
    \end{equation}
    
    \noindent
    The equation (\ref{formula: simplified_kinetic_equation}) requires boundary conditions to obtain a solution. In order to understand how the ISD distribution is modified inside the region of the supersonic solar wind, we assume the ISD flow is undisturbed out of the Termination Shock (TS) - the shock wave that restricts the region of the supersonic solar wind in the model of interaction between solar wind and interstellar medium. This helps to understand the modification of the ISD distribution inside the TS as opposed to that in the heliospheric interface (Alexashov et al. 2016). We consider the TS as a sphere with radius $r_{TS}$ and formulate the boundary condition as:
        
    \begin{equation} \label{formula: general_form_boundary}
        f_d({\bf r}, {\bf v})|_{r = r_{TS}, \: {\bf v} \cdot {\bf e}_{n} > 0} = f_{TS}({\bf v}),
    \end{equation}
    
    \noindent
    where $f_{TS}({\bf v})$ is the ISD distribution function on the TS and ${\bf e}_{n}$ is the interior unit normal to the sphere. Below we discuss the form of function $f_{TS}({\bf v})$ in more detail.
    
    \noindent
    To complete the correct mathematical formulation of the problem we should also set the boundary condition in the velocity space:
    
    \begin{equation} \label{formula: velocity_boundary}
        f_d({\bf r}, {\bf v})|_{v \rightarrow \infty} = 0.
    \end{equation}
    
    \noindent
    Note that each specific problem described by the formulation (\ref{formula: simplified_kinetic_equation}) - (\ref{formula: velocity_boundary}) is determined by a specific expression for the force term ${\bf F}({\bf r}, {\bf v})$ and for the boundary condition function $f_{TS}({\bf v})$.
        
    \subsection*{Force analysis}
    
    \begin{figure}
    	\centering
    	\includegraphics[width=11.5cm,height=8cm]{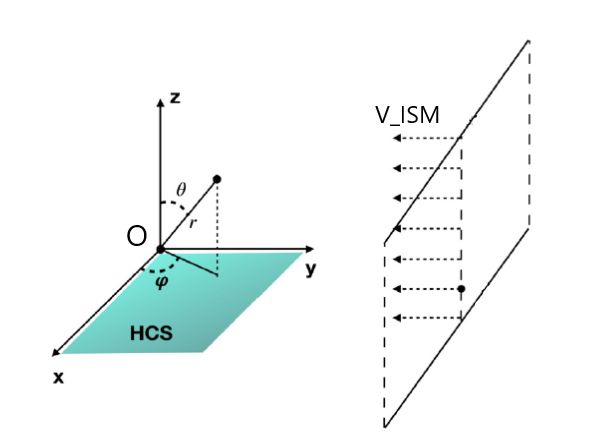}
    	\caption{The coordinate system. The Sun is located at O, the velocity of the LISM ${\bf v_{ISM}}$ is collinear to the $Oy$-axis. The $Oz$-axis coincides with the solar rotation axis. Spherical coordinates are introduced in the standard way.}
    	\label{picture: coord_system}
    \end{figure}
    
    \noindent
    Consider the Cartesian coordinate system as shown in Figure \ref{picture: coord_system}. Katushkina, Izmodenov (2019) provided the analysis of forces acting on the ISD particles. Four main forces act on the particles: the gravitational force ${\bf F}_{grav}$, the radiation pressure force ${\bf F}_{rad}$, the drag force ${\bf F}_{drag}$ due to interaction of dust grains with protons, electrons and neutrals and the electromagnetic force ${\bf F}_{el}$. Estimates show (Gustafson 1994) that in the heliosphere we can neglect the drag force.

    \noindent
    The expression for the gravitational force ${\bf F}_{grav}$ is:
    
    \begin{equation} \label{formula: gravitational_force}
        {\bf F}_{grav} = -\frac{G M_S}{r^2}{\bf e}_r,
    \end{equation}
    
    \noindent
    where $G$ is the gravitational constant, $M_S$ is the mass of the Sun. 
    
    \noindent
    Since ${\bf F}_{grav}$ is parallel ${\bf F}_{rad}$ and both are proportional to $r^{-2}$, it is convenient to introduce the parameter $\beta$:
    
    \begin{equation} \label{formula: beta_definition}
        \beta = \frac{|{\bf F}_{rad}|}{|{\bf F}_{grav}|}.
    \end{equation}
    
    \noindent 
    In this paper for the sake of simplicity we consider spherical particles. In this case $\beta$ depends only on the star characteristics and particle mass $m$ (see e.g. Katushkina, Izmodenov 2019). Here we use the $\beta = \beta(m)$ curve from Sterken et al. (2012) (green solid line in Figure 14). The resulting expression for the radiation pressure force is:
    
    \begin{equation} \label{formula: radiation_force}
        {\bf F}_{rad} = \beta \frac{G M_S}{r^2}{\bf e}_r.
    \end{equation}
    
    \noindent
    The magnetic field lines are frozen in the solar wind, that is why if we consider the reference frame related to the solar wind, one can derive the expression for the electromagnetic force using relative (with respect to the solar wind) velocity ${\bf v}_{rel}$ of dust particles:
    
    \begin{equation} \label{formula: electromagnetic_force}
        {\bf F}_{el} = \frac{q}{c_0 m_d} \left({\bf v}_{rel} \times {\bf B} \right),
    \end{equation}
    
    \noindent
    where ${\bf v}_{rel} = {\bf v} - {\bf v}_p$ is the dust particle velocity with respect to the solar wind, $q$ is the particle charge, $c_0$ is the speed of light, $m_d$ is the dust grain mass, ${\bf v}_p$ is the solar wind velocity, ${\bf B}$ is the solar magnetic field. The particle charge is expressed through the surface potential $U_d$ and the radius $a$: $q = U_d a$, and we consider $U_d$ constant in the supersonic solar wind (figure 2 from Alexashov et al. 2016, figure 2 from Slavin et al. 2012). Out of the region of the supersonic solar wind one should take account of the changes in value of the surface potential, but it is beyond the scope of the present work. The dust grain mass $m_d = \frac{4}{3}\rho_d\pi a^3$, where $\rho_d$ is the mass density of dust (here we consider astronomical silicates). We further assume uniform spherically symmetric solar wind: ${\bf v}_p = v_{sw}{\bf e}_r$, and for the solar magnetic field we use Parker's model:
        
    \begin{equation} \label{formula: magnetic_field}
    B_r = \pm B_E \left(\frac{r_E}{r}\right)^2, \:\: B_{\phi} = \mp \frac{B_E \Omega r_E}{v_{sw}}\left(\frac{r_E}{r}\right)\sin{\theta}, \:\: B_{\theta} = 0,
    \end{equation}
    
    \noindent
     where $B_E$ is the averaged solar magnetic field magnitude at the Earth's orbit, $r_E$ is the astronomical unit, $\Omega$ is the angular velocity of the solar rotation. The sign $\pm$ denotes the change in the polarity of the magnetic field across the heliospheric current sheet (HCS). Here for simplicity we assume a planar shape of the HCS (the plane $Oxy$ in Figure \ref{picture: coord_system}). In reality there is a non-zero angle between the solar rotation axis and the magnetic axis, so the HCS has the "ballerina skirt"\ shape. Here we also assume that the heliospheric magnetic field is stationary: the HCS plane is at rest and in the region $z > 0$ the magnetic field components $B_r < 0, \: R_{\varphi} > 0$ and vice versa. That is we ignore the 22-year solar cycle, which leads to polarity changes every 11 years accompanied by changes in the geometry of the HCS. In future we plan to expand our model to the time-dependent magnetic field case. 
    
    \noindent
    Thus, the expression for ${\bf F}({\bf r}, {\bf v})$ is:
    
    \begin{equation} \label{formula: result_force}
        {\bf F}({\bf r}, {\bf v}) = (\beta - 1)GM_s \cdot \frac{{\bf e}_r}{r^2} + \frac{q}{c_0m_d}(({\bf v} - {\bf v}_p) \times {\bf B}).
    \end{equation}
    
    \subsection*{Boundary condition}
    
    \noindent
    Let us assume that in the LISM there is a flux of ISD particles with the average velocity ${\bf v}_{ISM}$ and dispersion of the $v_z$ velocity component. The interstellar magnetic field has spatial and temporal inhomogeneities which act as sources for the acceleration of ISD particles (Hoang et al. 2012). This acceleration is the reason for variations in the velocities of individual dust particles and, therefore, appearance of dispersion in the ISD velocity distribution.  Below we demonstrate that relatively small values of dispersion of the $v_z$ component significantly influence the results. Then the expression for $f_{TS}(\bf v)$ is:
        
    \begin{equation} \label{formula: boundary_condition}
        f_{TS}({\bf v}) = n_{ISM} \delta(v_x) \delta(v_y + v_{ISM}) \frac{1}{\sigma_z \sqrt{2\pi}}\exp{\left(-\frac{v_z^2}{2\sigma_z^2}\right)},
    \end{equation}
    
    \noindent
    where $\delta$ is the Dirac delta-function, $\sigma_z$ is the dispersion of the $v_z$ component. As $\sigma_z \rightarrow 0$ the expression (\ref{formula: boundary_condition}) degenerates into the singular distribution function for the case when all dust particles have the same velocities ${\bf v}_{ISM}$:
    
    \begin{equation} \label{formula: boundary_condition_identical}
        f_{TS}({\bf v}) = n_{ISM} \delta(v_x) \delta(v_y + v_{ISM}) \delta(v_z),
    \end{equation}
    
    \noindent
    and the formulation of the problem is identical to the one given by Mishchenko et al. (2020). 
    
    \noindent
    Slavin et al. (2012) take account of the dispersion by addition of the supplementary velocity component lying in the plane perpendicular to the direction of the interstellar magnetic field. This supplementary velocity component has constant absolute value (3 km/s) and random direction in the above-mentioned plane. In the present work we model the dispersion of the $v_z$ component using the normal distribution.
    
    \subsection*{Dimensionless formulation of the problem}
    
    \noindent
    As a characteristic distance we consider $L_1 = \frac{GM_S}{v_{ISM}^2}$ and as a characteristic velocity -- $v_{ISM}$. Since the problem is linear and homogeneous in $f_d({\bf r}, {\bf v})$ we can substitute $f_d \rightarrow \frac{f_d}{n_{ISM}}$ and eliminate $n_{ISM}$ in (\ref{formula: boundary_condition}). The dimensionless formulation of the problem (\ref{formula: simplified_kinetic_equation}) - (\ref{formula: velocity_boundary}), (\ref{formula: result_force}), (\ref{formula: boundary_condition}) is:
    
    \begin{equation} \label{formula: dimensionless_formulation}
        \begin{cases}
            \hat{{\bf v}} \cdot \frac{\partial \hat{f}_d}{\partial \hat{{\bf r}}} + \hat{{\bf F}} \cdot \frac{\partial \hat{f}_d}{\partial \hat{{\bf v}}} = 0, \\
            \hat{f}_d(\hat{{\bf r}}, \hat{{\bf v}})|_{\hat{r} = \hat{r}_{TS}, \: \hat{{\bf v}} \cdot {\bf e}_n > 0} = \delta(\hat{v}_x)\delta(\hat{v}_y + 1)\frac{1}{\hat{\sigma}_z\sqrt{2\pi}}\exp{\left(-\frac{\hat{v}^2_z}{2\hat{\sigma}_z^2}\right)}, \\
            \hat{f}_d(\hat{{\bf r}}, \hat{{\bf v}})|_{\hat{v} \rightarrow \infty} = 0 ,
        \end{cases}
    \end{equation}
    
    \noindent
    where $\hat{{\bf r}} = \frac{{\bf r}}{L_1}, \: \hat{{\bf v}} = \frac{{\bf v}}{v_{ISM}}, \: \hat{f}_d = \frac{f_d}{v_{ISM}^3}, \: \hat{{\bf F}} = \frac{{\bf F} L_1}{v_{ISM}^2}, \: \hat{\sigma}_z = \frac{\sigma_z}{v_{ISM}}, \hat{r}_{TS} = \frac{r_{TS}}{L_1}$. The expression for the sum of forces (\ref{formula: result_force}) in the dimensionless form is:
    
    \begin{equation} \label{formula: dimensionless_force}
        \hat{{\bf F}}= (\beta - 1) \frac{{\bf e}_r}{\hat{r}^2} + sgn(\hat{z}) \frac{v_{em}^2}{v_{ISM}^2} \left(\frac{v_{ISM}}{v_{sw}}\hat{{\bf v}} - {\bf e}_r\right) \times \left(-\frac{L_{\Omega}}{L_1} \frac{{\bf e}_r}{\hat{r}^2} + \frac{sin \theta}{\hat{r}} {\bf e}_{\varphi}\right),
    \end{equation}
    
    \noindent
    with $L_{\Omega} = \frac{v_{sw}}{\Omega}, \: v_{em}^2 = \frac{q B_E \Omega r_E^2}{c_0 m_d}$. The dimensionless formulation of the problem contains five dimensionless parameters:
    
    \begin{equation} \label{formula: full_dimensionless_parameters}
    \hat{\sigma}_z,\: \beta,\: \varepsilon = \frac{v_{em}^2}{v_{ISM}^2} = \frac{3 U_d B_E R_E^2 \Omega}{4 \pi c_0 \rho_d a^2 v_{ISM}^2},\: \frac{v_{ISM}}{v_{sw}}, \: \frac{L_{\Omega}}{L_1}.
    \end{equation}
    
    \subsection*{Trajectories in the plane of symmetry}
    
    \noindent
    Let us consider the projection of the force $\hat{{\bf F}}$ on the $x$-axis:
    
    \begin{equation} \label{formula: x_axis_projection}
        \hat{F}_x = (\beta - 1)\frac{\hat{x}}{\hat{r}^3} + sgn(\hat{z})\varepsilon \frac{\hat{x}\hat{z}}{\hat{r}^3} + 
        sgn(\hat{z})\varepsilon\frac{v_{ISM}}{v_{sw}}\left(-\frac{L_{\Omega}}{L_1}\frac{\hat{v}_y\hat{z} - \hat{v}_z\hat{y}}{\hat{r}^3} - \frac{\hat{v}_z\hat{x}}{\hat{r}^2}\right).
    \end{equation}
    
    \noindent
    Since $\frac{v_{ISM}}{v_{sw}} \approx 0.05 << 1$, one can neglect the corresponding term, and the expression for $\hat{F}_x$ in a simplified form is:
    
    \begin{equation} \label{formula: simplified_projection}
        \hat{F}_x = (\beta - 1)\frac{\hat{x}}{\hat{r}^3} + sqn(\hat{z})\varepsilon \frac{\hat{x}\hat{z}}{\hat{r}^3}.
    \end{equation}
    
    \noindent
    Therefore, dust particles with initial parameters:
    
    \begin{equation} \label{formula: flat_particle_conditions}
        \hat{x}|_{TS} = 0, \:\:\: \hat{v}_x|_{TS} = 0,
    \end{equation}
    
    \noindent
    can't leave the plane $\hat{x} = 0$ under the action of the force (\ref{formula: simplified_projection}) according to Picard's existence and uniqueness theorem. For simplicity we consider only such trajectories in this article.
    
    \subsection*{Monte-Carlo approach}
    
	\begin{figure}
	    \centering
         \includegraphics[width=10cm,height=9cm]{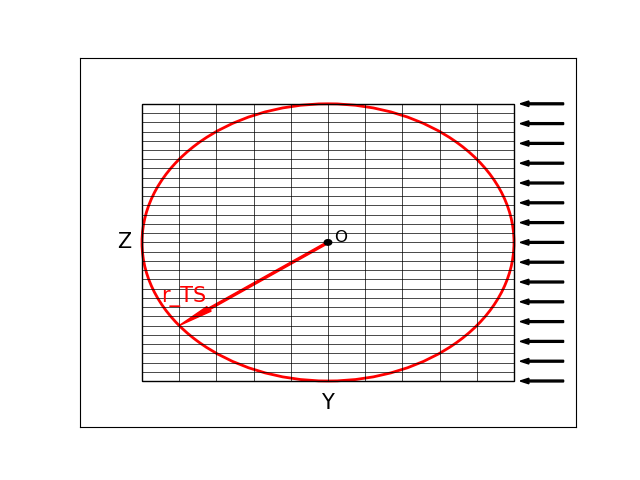}
	     \caption{The computational domain is a square with the side $2 \hat{r}_{TS} = \frac{r_{TS}}{L_1}$ which is divided into rectangular cells $\Delta \hat{y} \times \Delta \hat{z}$.}
	     \label{picture: grid}
    \end{figure}    
    
    \noindent
    To solve the kinetic equation we use the Monte-Carlo method. The computational domain is divided into rectangular cells $\Delta \hat{y} \times \Delta \hat{z}$ (Figure \ref{picture: grid}) and $\Delta \hat{z} << \Delta \hat{y}$ because the singular layers found in Mishchenko et al. (2020) are oriented horizontally. Moreover, since their thickness approaches zero one should decrease $\Delta \hat{z}$ in order to detect these peculiarities by the Monte-Carlo modeling.
    
    \noindent
    For a dust particle we generate randomly its initial velocity and location on the sphere with radius $r = r_{TS}$ according to the distribution function $f_{TS}({\bf v})$ from (\ref{formula: dimensionless_formulation}). During the motion of the particle in the heliosphere we record the time $t_i$ of the particle in the computational domain cells ($t_i = 0$ if particle does not cross the corresponding cell). Then, by the definition of the distribution function and number density, and the law of large numbers we have:
        
    \begin{equation} \label{formula: distribution_function_m-c}
        \frac{\hat{F}_0}{N}\sum_{i=1}^N \frac{t_i}{\Delta \hat{{\bf r}}_c \Delta \hat{{\bf v}}_c} \rightarrow f_d(\hat{{\bf r}}_c, \hat{{\bf v}}_c)
    \end{equation}
    
    \begin{equation} \label{formula: density_m-c}
        \frac{\hat{F}_0}{N}\sum_{i=1}^N \frac{t_i}{\Delta \hat{{\bf r}}_c} \rightarrow n_d(\hat{{\bf r}}_c),
    \end{equation}
    
    \noindent
    where $N$ is the number of particles, $\Delta {\bf r}_c \Delta {\bf v}_c$ is the cell volume in the phase space, $\hat{F}_0$ is the flux of the dust particles through the outer surface per unit of time in the dimensionless form:
    
    \begin{equation} \label{formula: start_flux}
        \hat{F}_0 = \int\limits_{-\frac{\pi}{2}}^{\frac{\pi}{2}}\left(\int\limits_{\left(\hat{{\bf v}} \cdot {\bf e}_n\right) > 0} \left(\hat{{\bf v}} \cdot {\bf e}_n\right) \hat{f}_{TS}(\hat{{\bf v}}) d\hat{{\bf v}}\right)\hat{r}_{TS}d\varphi = 2\hat{r}_{TS}
    \end{equation}
    
    \subsection*{Technical characteristics}
    
    \noindent
    In this paper we consider particles with the radius $a = 0.37 \mu m$. For these particles $\beta = 1$ and consequently the gravitational and radiation pressure forces cancel out in (\ref{formula: dimensionless_force}).
    
    \noindent
    For computations we use the following values of the parameters: $r_{TS} = 100 \:\:$a.u, $v_{ISM} = 26.4 \:\:$km/s, $M_S = 2 \cdot 10^{30} \:\: \text{kg}$, $v_{sw} = 400 \:\:$km/s, $\Omega = 2.9 \cdot 10^{-6} \:\:$1/s, $U_d = +3 \:\:$V, $B_E = 30 \mu$G, $R_E = 1 \:\:$a.u., $\rho_d = 2500 \:\:$kg/m$^{3}$.

    \noindent
    For all figures with results in this paper, unless otherwise specified, the cell size inside the computational domain is $0.1 \: a.u. \: \times \: 0.001 \: a.u.$ in the $Oy$-  and $Oz$-directions, respectively. To solve the system of ODEs for the trajectory of a particle the fourth order Runge-Kutta method was used.
    
    \noindent
    Note that the selected fixed location of the HCS corresponds to the case when all ISD particles are attracted to the HCS (focusing phase). In order to understand it let us consider the $z$-axis projection of (\ref{formula: dimensionless_force}):
        
    \begin{equation}
        \hat{F}_z = (\beta - 1) \frac{\hat{z}}{\hat{r}^3} + sgn(\hat{z})\varepsilon \left(- \frac{\hat{x}^2 + \hat{y}^2}{\hat{r}^3} + \frac{v_{ISM}}{v_{sw}}\left(-\frac{L_{\Omega}}{L_1}\frac{\hat{v}_x \hat{y} - \hat{v}_y \hat{x}}{\hat{r}^3} + \frac{\hat{v}_x \hat{x} + \hat{v}_y \hat{y}}{\hat{r}^2}\right)\right),
    \end{equation}
    
    \noindent
    where again $\frac{v_{ISM}}{v_{sw}} \approx 0.05 << 1$, that is why at large heliospheric distances the leading term is: 
        
    \begin{equation}
        -sgn(\hat{z}) \varepsilon \frac{\hat{x}^2 + \hat{y}^2}{\hat{r}^3}.
    \end{equation}
    
    \noindent
    It is seen, that in the case $\hat{z} > 0$ the $z$-axis component $\hat{F}_z < 0$, so the ISD particles are attracted to the HCS. In the case $\hat{z} < 0$ we have $\hat{F}_z > 0$
    
    \section*{RESULTS}
    
    \subsection*{Singularities in density}
    
	\begin{figure}
		\centering
		\includegraphics[width=15cm,height=8.5cm]{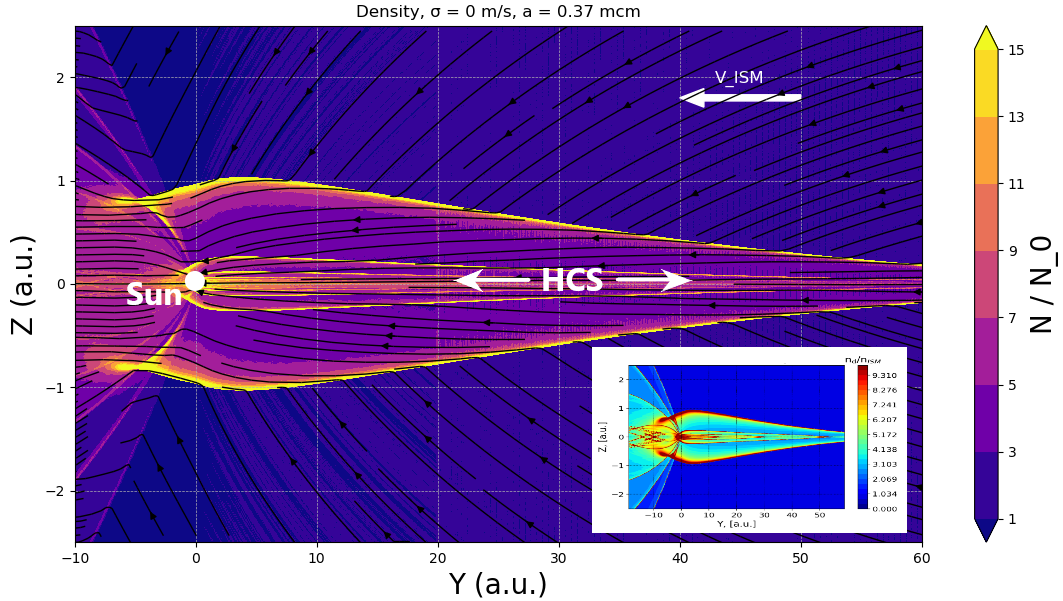}
		\caption{Map of the density distribution in the plane $X = 0$ in the case without dispersion. Yellow color envelopes correspond to caustics. Relative statistical error is limited by 2-3 \% at each point. The number of trajectories $N = 2000000$. For the sake of comparison, the panel at the right bottom presents the results of Mishchenko et al. (2020) obtained for the same conditions. The radius of particles is 0.37 $\mu$m.}
		\label{picture: full_density_a=0_37}
	\end{figure}

	\begin{figure}
		\centering
		\includegraphics[width=13cm,height=6.2cm]{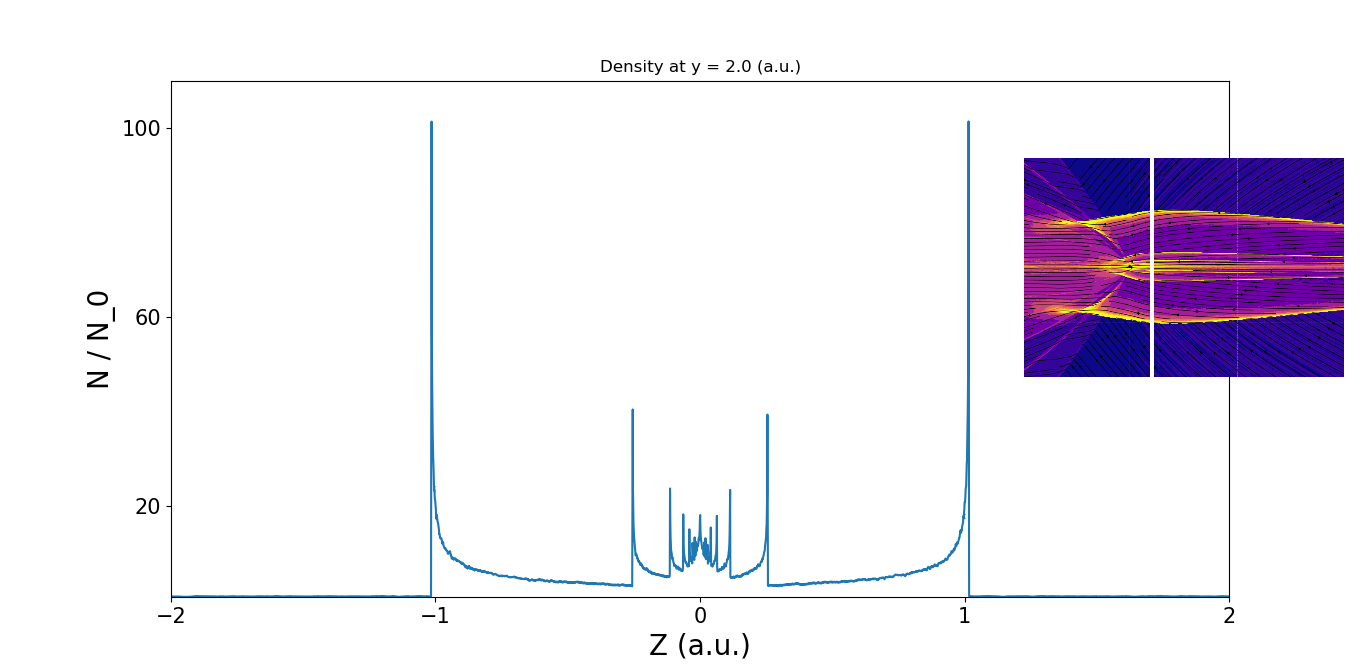}
		\caption{The distribution of ISD density along the line ($X = 0, Y = 2$). The cell size is $0.1 \:\:$a.u. $\times \: 0.001 \:\:$a.u. Relative statistical error is limited by 2-3 \% at each point. The number of trajectories $N = 2000000$. The radius of particles is 0.37 $\mu$m.}
		\label{picture: sech_y=2_a=0_37}
	\end{figure}

	\begin{figure}
		\centering
		\includegraphics[width=14cm,height=7cm]{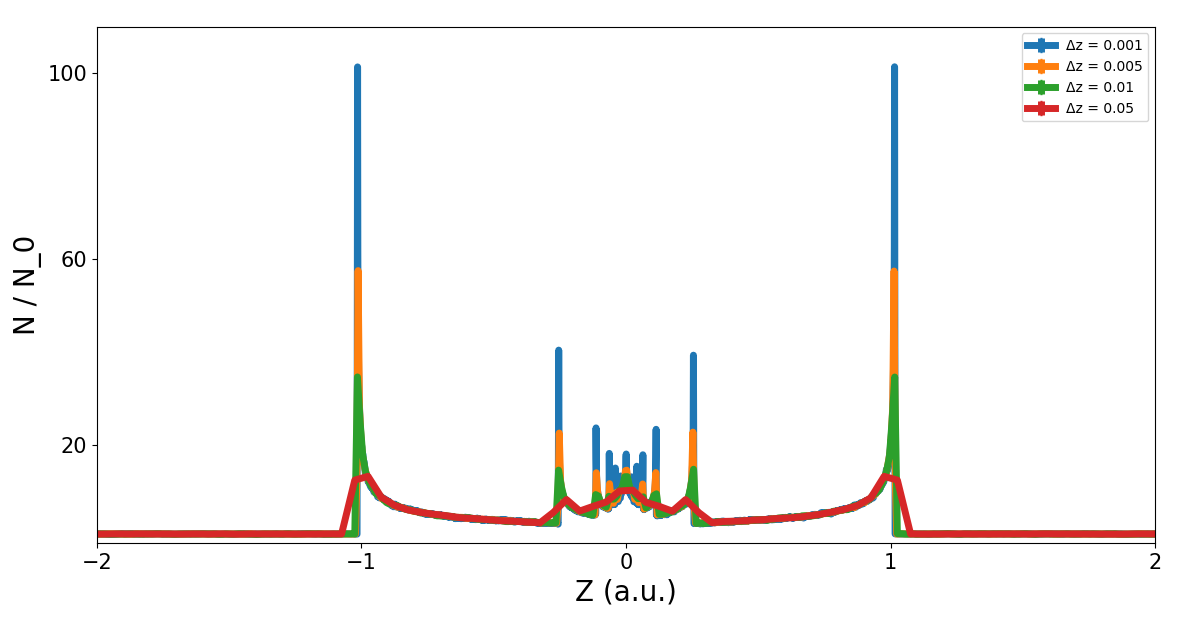}
		\caption{The distribution of ISD density along the same line as in Figure \ref{picture: sech_y=2_a=0_37}. The four lines of the  different colour correspond to the different cell sizes in the $z$-direction ($\Delta y = 0.1 \:\:$a.u., $\Delta z = \{0.05 \:\:$a.u., $0.01 \:\:$a.u., $0.005 \:\:$a.u., $0.001 \:\:$ a.u.$\}$). Relative statistical error is limited by 2-3 \% at each point. The number of trajectories $N = 2000000$. The radius of particles is 0.37 $\mu$m.}
		\label{picture: cell_comparison_a=0_37}
	\end{figure}

	\begin{figure}
		\centering
		\includegraphics[width=16cm,height=7.5cm]{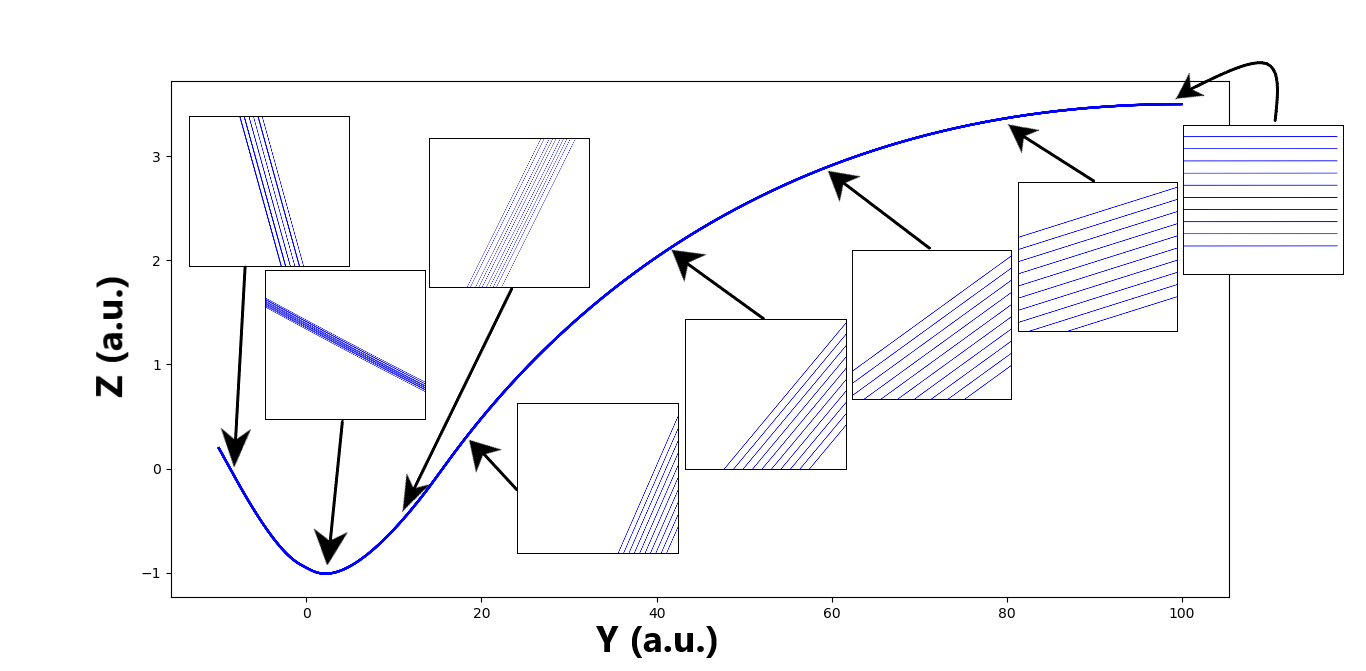}
		\caption{The tube of trajectories from a small region on the outer boundary. The tube is compressed up to a thousand times at small heliocentric distances. The point of minimal width corresponds to a point on the caustic. Small boxes demonstrate trajectories view at a large scale. The box sizes are $0.6 \:\:$a.u. $\times \: 0.025 \:\:$a.u.  The radius of particles is 0.37 $\mu$m.}
		\label{picture: combined_tracks_new}
	\end{figure}    
    
    \noindent
    It was shown by Mishchenko et al. (2020) that in the case of zero dispersion the ISD trajectories form \textit{caustics} at which the number density of ISD is infinite. A caustic is an envelope of the ISD trajectories. By definition, every segment of a caustic is tangent to an infinite number of the ISD trajectories, that is the reason for density singularities origin. The distribution of the dust density has multiple singularities. This result was obtained by the Lagrangian approach. In this Section we demonstrate that the singularities can be also obtained by the Monte-Carlo approach (although, perhaps, at the cost of computational efficiency).
    
    \noindent
    Figure \ref{picture: full_density_a=0_37} shows the map of the ISD density distribution as well as the ISD streamlines. The map shows the region in the vicinity of the HCS. Symmetrical yellow lines in this Figure are the above-mentioned caustics. In the case of the Monte-Carlo simulation they represent the thin regions where a sharp density peak is found (Figure \ref{picture: sech_y=2_a=0_37}). Inside the area delimited by the caustics there is a complex structure of the ISD density distribution with many local peaks. 
    
    \noindent
    Since the computational domain consists of \textit{finite} size cells, high spatial resolution of the numerical grid is required to detect the density singularities with high precision by the Monte-Carlo modeling. Figure \ref{picture: cell_comparison_a=0_37} shows how the ISD density distribution along the line ($X = 0, Y = 2$) changes with variation of cell size $\Delta z$. The ISD density at cells containing caustic points increases with decreasing $\Delta z$ and, therefore, the ISD density singularities are located at these cells.
    
    \noindent
    A simple explanation for the formation of the caustics is as follows. Let us look at the particles originating from a small region on the boundary of the computational domain (Figure \ref{picture: combined_tracks_new}). This flux tube is compressed with decreasing heliocentric distance and reaches its minimal width (approaching zero) exactly at the caustic points. Considering the conservation of mass for a flux tube:
    
    \begin{equation*}
        n_1 v_1 \Sigma_1  = n_2 v_2 \Sigma_2   
    \end{equation*}
    
    \noindent
    a minimal value of tube width $\Sigma$ corresponds to the maximal value of the density $n$, because the value of the $y$-component velocity $v$ is approximately constant.
        
    \subsection*{Effects of velocity dispersion}
    
	\begin{figure}
	    \centering
	    \includegraphics[width=14cm,height=7.5cm]{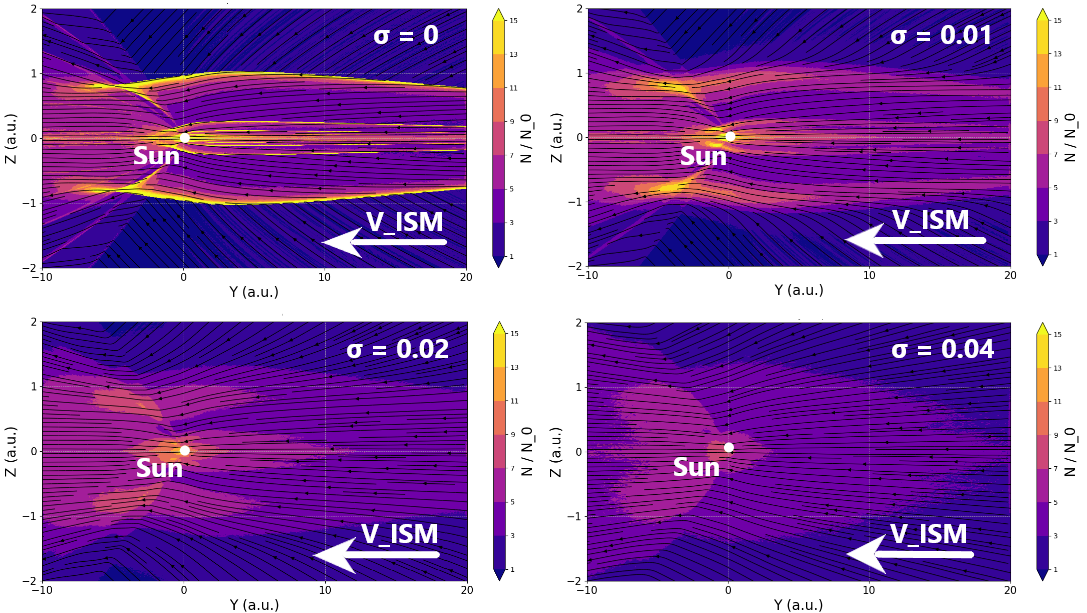}
        \caption{Comparison of the maps of the ISD density distributions with different dispersion $\hat{\sigma}_z: 0, 0.01, 0.02, 0.04$. With increasing dispersion $\hat{\sigma}_z$ the caustics are smeared and the density singularities disappear. Relative statistical error is limited by 2-3 \% at each point. The number of trajectories $N = 2000000$. The radius of particles is 0.37 $\mu$m.}
       \label{picture: map_comparison_a=0_37}
    \end{figure}

    \begin{figure}
	    \centering
        \includegraphics[width=14cm,height=7cm]{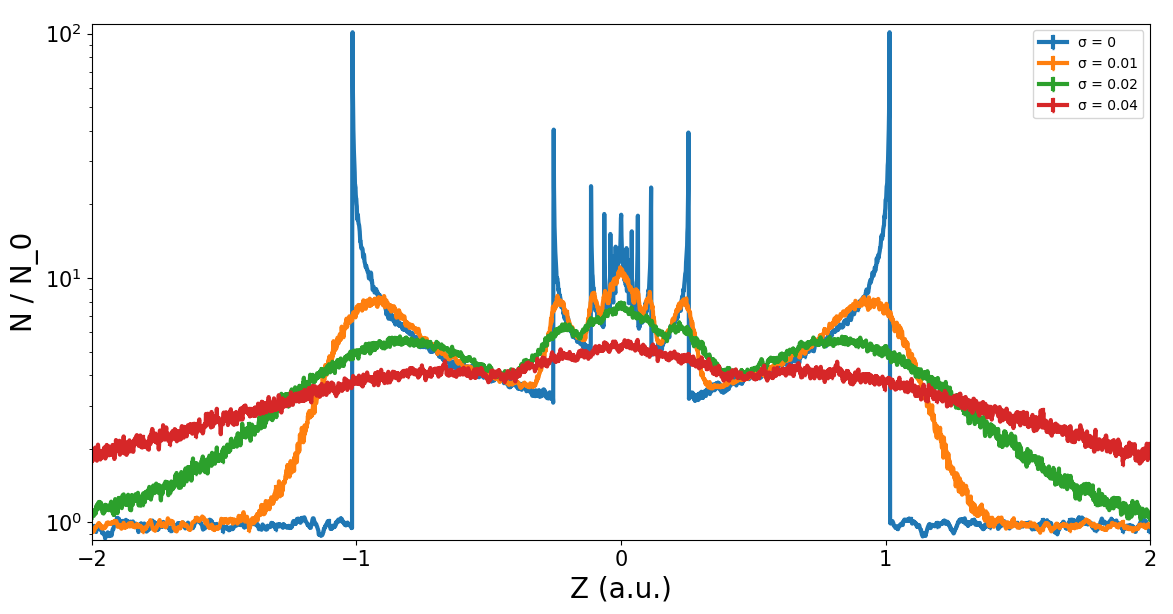}
	    \caption{Comparison of the ISD density distributions along the line ($X = 0, Y = 2$) with different dispersion $\hat{\sigma}_z: 0, 0.01, 0.02, 0.04$. The structure of the density distribution changes drastically with variation of dispersion. Relative statistical error is limited by 2-3 \% at each point. Th number of trajectories $N = 2000000$. The radius of particles is 0.37 $\mu$m.}
	    \label{picture: sech_comparison_a=0_37}
    \end{figure}

    \begin{figure}
	    \centering
	    \includegraphics[width=14cm,height=7cm]{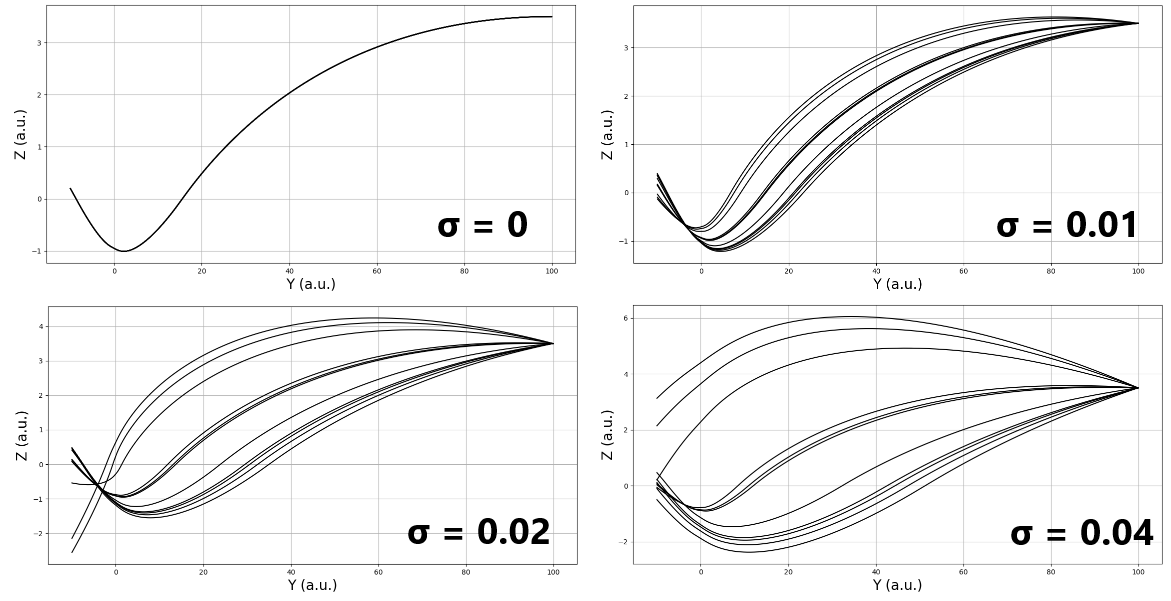}
        \caption{Trajectories of particles originating from a small region on the boundary of the computational domain for different values of $\hat{\sigma}_z$. Particle trajectories are scattered for non-zero dispersion, and no singularities appear. The radius of particles is 0.37 $\mu$m.}
	    \label{picture: tracks_comparison}
    \end{figure}

    \begin{figure}
	    \centering
	    \includegraphics[width=14cm,height=7cm]{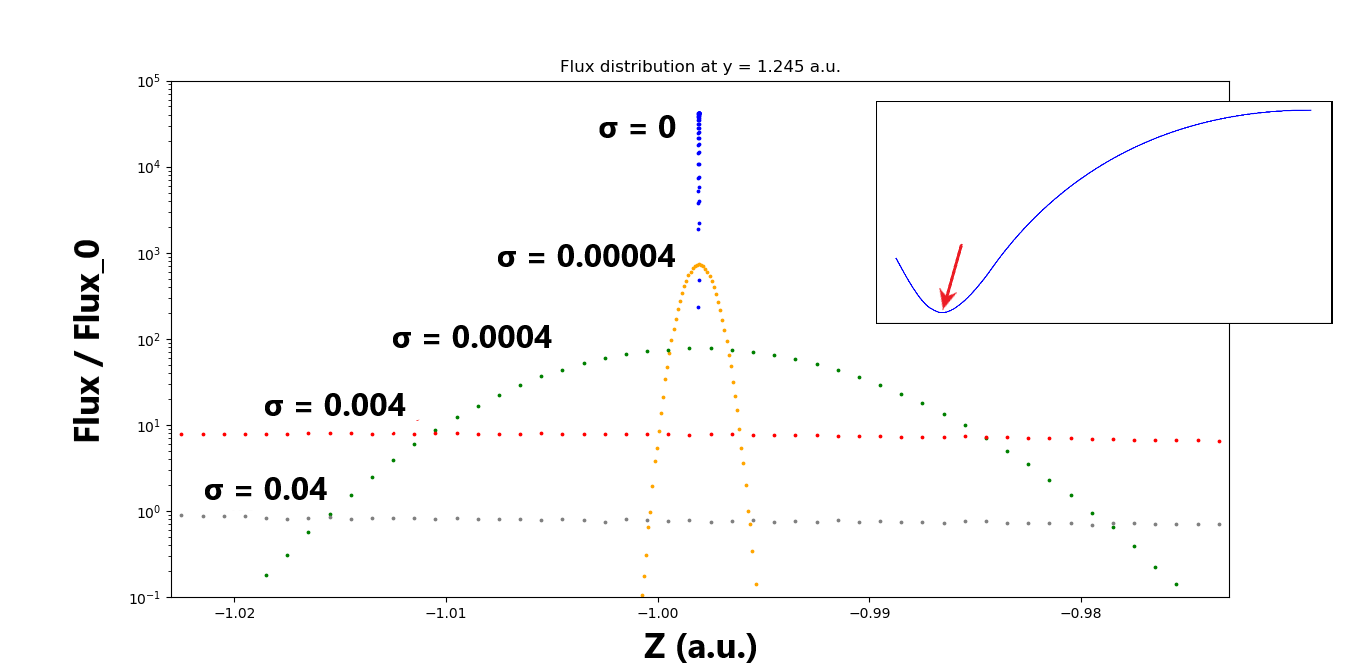}
	    \caption{Comparison of the fluxes carried by the ISD trajectories originating from a small region on the boundary of the computational domain with different dispersion $\hat{\sigma}_z: 0, 0.00004, 0.0004, 0.004, 0.04$. For each computation we have the convergence in cell size, Runge-Kutta integration steps and the number of simulated particles. The maximal value of fluxes is approximately inversely proportional to the dispersion $\hat{\sigma}_z$. For computation without dispersion the cell size in the $z$-axis direction is $10^{-6} \:\:$(a.u.). Relative statistical error is limited by 5 \% at each point. The radius of particles is 0.37 $\mu$m.}
	    \label{picture: flux_comparison}
    \end{figure}    

\begin{figure}
    	\centering
    	\includegraphics[width=14cm,height=7.5cm]{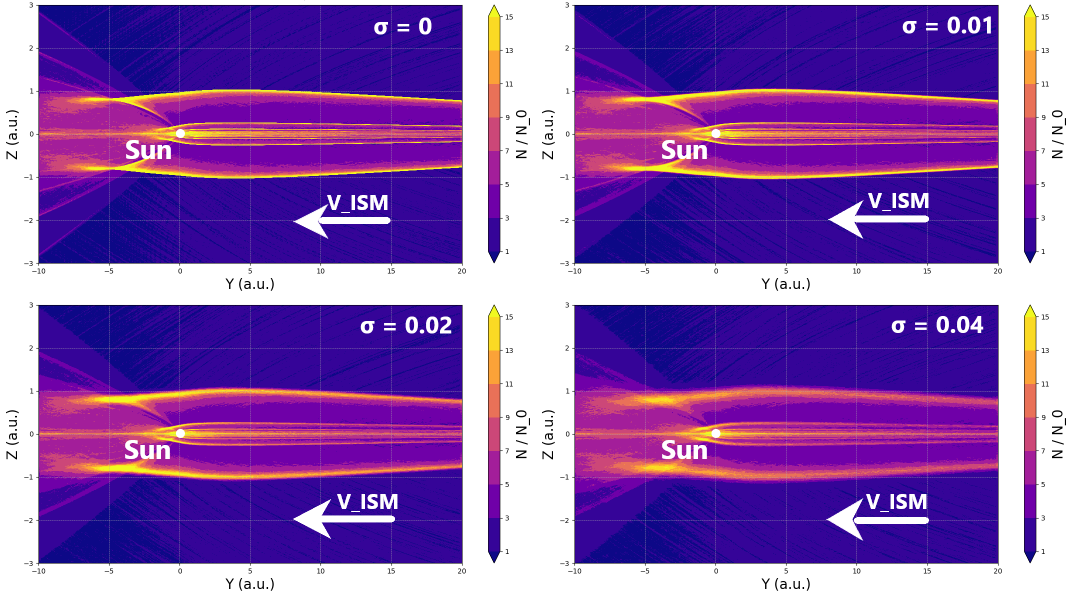}
    	\caption{Comparison of density map distributions with different dispersion of the $v_y$ velocity component: $0, 0.01, 0.02, 0.04$. The shape of the overdensity regions remains virtually unchanged with the variation in the dispersion. Relative statistical error is limited by 5-6 \% at each point. Number of trajectories $N = 200000$. Radius of particles is 0.37 $\mu$m.}
    	\label{picture: v_y_comparison}
    \end{figure}

    \noindent
    In this paper we mainly consider the dispersion of the $v_z$ component. We do not consider the dispersion of the $v_x$ component at all because we only study the plane of symmetry $X = 0$, and Figure \ref{picture: v_y_comparison} shows that the dispersion of the $v_y$ component has less impact on the density distribution than the dispersion of the $v_z$ component.

    \noindent
    To explore the effect of velocity dispersion we performed the calculations for the set of $\hat{\sigma}_z$ values: $0, 0.01, 0.02, 0.04$. Figure \ref{picture: map_comparison_a=0_37} presents the density maps obtained for the four values of $\hat{\sigma}_z$. With increasing $\hat{\sigma}_z$ the density maxima are smeared and their singularities disappear. The regions of overdensity remain only in the vicinity of the HCS. This clumping up of dust particles is associated with an increase in the magnitude of the Lorentz force at small heliocentric distances, which leads to decrease in the amplitude of the particle oscillations around the HCS. Then, the gross tendency is for the ISD to converge to the HCS plane and so the regions of overdensity appear. In Figure \ref{picture: sech_comparison_a=0_37} we can see how the density at cells containing caustic points changes quantitatively with variation of $\hat{\sigma}_z$. Small values of dispersion in the boundary velocity distribution drastically change the density distribution in the heliosphere. 
    
    \noindent
    The reason for the disappearance of the singularities is clearly seen from Figures \ref{picture: tracks_comparison}, \ref{picture: flux_comparison}. Figure \ref{picture: tracks_comparison} shows trajectories of particles originating from a small region on the outer boundary of the  computational domain for different $\hat{\sigma}_z$ values. As it was mentioned above, singularities appear where the width of the flux tube approaches zero. Particle trajectories are scattered  for non-zero dispersion, and therefore no singularities appear. Density flux distributions in the vicinity of the point corresponding to the caustic by particles originating from a small region on the outer boundary are demonstrated in Figure \ref{picture: flux_comparison} for different $\hat{\sigma}_z$ values. One can appreciate that the maxima of these density flux distributions are approximately inversely proportional to the value of the dispersion $\hat{\sigma}_z$. Thus even extremely small values of dispersion drastically influence the ISD distribution inside the heliosphere.
           
    \noindent
    In order to study the influence of the dispersion of the $v_y$ component on the density distribution instead of the expression (\ref{formula: boundary_condition}) we should use the following boundary condition function:
        
    \begin{equation} \label{formula: v_y_boundary_condition}
        f_{TS}({\bf v}) = n_{ISM} \delta(v_x) \frac{1}{\sigma_y \sqrt{2\pi}}\exp{\left(-\frac{(v_y + v_{ISM})^2}{2\sigma_y^2}\right)} \delta(v_z),
    \end{equation}
    
    \noindent
    which in the dimensionless form is:
    
    \begin{equation} \label{formula: dimensionless_v_y_boundary_condition}
        f_{TS}(\hat{{\bf v}}) = \delta(\hat{v}_x) \frac{1}{\hat{\sigma}_y \sqrt{2\pi}}\exp{\left(-\frac{(\hat{v}_y + 1)^2}{2\hat{\sigma}_y^2}\right)} \delta(\hat{v}_z).
    \end{equation}
    
    \noindent
    Figure \ref{picture: v_y_comparison} shows the comparison of density distributions for the cases with different values of dispersion $\hat{\sigma}_y$ of the $v_y$ component ($\hat{\sigma}_y$: $0, 0.01, 0.02, 0.04$). We can see that the shape of the overdensity region has remained virtually unchanged for the chosen dispersion values. Since the same dispersion values were used previously for the $v_z$ component, one can conclude that the dispersion of the $v_z$ component has greater impact on the density distribution than the dispersion of the $v_y$ velocity component. This is because the regions of overdensity are stretched along the $Oy$-axis and, therefore, in the stationary case small variations in the $v_y$ component can't significantly influence the ISD density distribution.

    \section*{CONCLUSION}
    
    \noindent
    In this paper we demonstrated that the singularities of the ISD density in the heliosphere, discovered using the Lagrangian approach in Mishchenko et al. (2020), can also be found by the Monte-Carlo simulations. This requires super-small computational cells. In our calculations the required size of a cell (in the $z$-direction ) is $10^{-3} \:\:$ a.u. Having a such size of the cells in the whole domain is computationally unrealistic. Weaker resolution (i.e. larger cells) does not allow to find the caustics.

    \noindent
    Dispersion was introduced as a normal distribution of one of the velocity component. It was shown that the density singularities are smeared due to dispersion. The regions of overdensity are smoothed and remain only in the vicinity of the heliospheric current sheet. It is known (Hoang et al. 2012) that the velocity dispersion can reach values of approximately 15 \% due to spatial and temporal inhomogeneities in the interstellar magnetic field. Significant qualitative and quantitative changes in the density distribution emerge even for 5 \% dispersion as it was shown. Thus, the velocity dispersion is an extremely important effect that strongly influences the ISD density distribution inside the heliosphere.
    
    \noindent
    In the future we plan to develop our model to the case of the time-dependent solar magnetic field in accordance with the 22-year solar cycle (in this paper we considered the solar magnetic field just in one focusing phase (Mann 2010)). Certainly this is a highly important effect that has a major impact on the ISD density inside the heliosphere and which is necessary to take into consideration.
        
    \section*{Acknowledgements}
    
    \noindent
    The authors are grateful to the Government of Russian Federation and the Ministry of Science and Higher Education for the support by grant 075-15-2020-780 (N13.1902.21.0039). We thank D. B. Alexashov, I. Baliukin and A. Granovskiy for useful discussions and for the help with preparation of the manuscript. This work is supported by grant 18-1-1-22-1 of the "Basis"\ Foundation.

\end{document}